# Predicted Performance Bounds of Thermochromism Assisted Photon Transport for Efficient Solar Thermal Energy Storage

Inderpreet Singh[a] and Vikrant Khullar[b,*]
[a]Mechanical Engineering Department, Chandigarh University, Gharuan, Mohali-140413, Punjab, India
[b]Mechanical Engineering Department, Thapar Institute of Engineering & Technology, Patiala-147004, Punjab, India

*Corresponding author. Email address: vikrant.khullar@thapar.edu

**ABSTRACT**
Efficient storage of solar thermal energy is still one of the major bottlenecks in realizing dispatchable solar thermal systems. Present work is a significant step in this direction, wherein, we propose, thermochromism assisted photon transport based optical charging for efficient latent heat storage. Seeding thermochromic nanoparticles into the phase change material (PCM) allows for dynamic control of PCM's optical properties - aiding deeper penetration of photons and hence significantly enhancing the photon-nanoparticle interactions. Moreover, carefully tailoring of transition temperature near the melting temperature allows for efficient non-radiative decay of the absorbed photon energy and that too under nearly thermostatic conditions. In particular, the present work serves to develop a mechanistic opto-thermal theoretical modelling framework to compute melting front progression, latent heat storage and sensible heat discharging capacities pertinent to thermochromism assisted photon transport. Moreover, to truly assess and quantify the benefits of the aforementioned charging route, a host of other possible charging routes (viz., thermal and non-thermochromic optical charging) have also been dealt with. Detailed analysis reveals that relative to the thermal charging route, thermochromism assisted optical charging offers significant enhancements in terms of melting front progression (approximately 152%) and latent heat storage capacity (approximately 167%). Overall, thermochromism assisted photon transport is a synergistic approach which allows for simultaneous collection and storage of solar energy at accelerated rates without requiring the PCM to be heated to high temperatures.

## 1. INTRODUCTION

At present, 90% of the primary energy generated is utilized or dissipated in the form of heat [1]. Potentially, to cater to this demand, the electromagnetic energy from the sun could be converted into useful thermal energy through photo-thermal energy conversion process. However, solar energy resource is highly intermittent; therefore, to ensure reliable and dispatchable supply, it necessitates coupling of solar-thermal conversion platforms with efficient solar energy storage systems. To this end, thermal energy storage (TES), particularly, in the form of latent heat of phase change materials (PCMs) has been shown to be a simple, safe, and high energy density storage route. Invariably, latent heat storage involves thermal charging of the PCM, wherein, thermal energy transfer to the PCM happens via a thermally conducting surface. However, owing to inherently low thermal conductivity of the PCM, the charging process is slow and inefficient [2, 3]. To improve the charging rates, researchers have employed thermally conducting nano-sized structures (as additives) and or macroscopic matrices to improve the overall thermal conductivity of the PCMs [4-11] The aforementioned approaches could essentially be categorized as "thermal or surface charging" of PCM characterized by "conduction" as the predominant mode of heat transfer and which is often assisted by convection transport [12].

More recently, an altogether different approach; viz., instead of thermal or surface charging of the PCM, the concept of "optical or volumetric charging" has been explored. Herein, the nanoparticles laden PCM (nano-PCM) directly interacts with the incident solar energy and owing to the photon-nanoparticle interactions, solar energy conversion into the latent heat of the PCM is highly efficient. At optimum nanoparticles concentration, significant enhancements in charging rates have been reported [3, 13]. However, optical charging could only be engineered up to a certain depth of PCM owing to finite penetration depth of the incident solar energy. After the aforementioned depth, the heat transfer to the subsequent layers happens predominantly through conduction (and in certain cases assisted by natural convection). Moreover, once a particular layer of nano-PCM has been melted, it acts as an optical barrier for the incident sunlight to reach the subsequent layers.

Very recently, to address this concern, researchers have devised dynamic tuning of nano-PCM optical properties to increase the penetration depth of the incident solar energy. In particular, external magnetic field has been employed to move the magnetic nanoparticles away from path of the incident sunlight in the melted portion. Thus, allowing the incident light to always strike the "fresh" solid nano-PCM layer [13-15]. Although, the aforementioned approach has shown to considerably enhance the melting rate; it is energy intensive and requires very careful real-time control - thus, difficult to practically implement in real systems.

Unequivocally, the optical charging approach has shown promising results; it is relatively new and warrants further investigation to realize systems which could be deployed in real world conditions.

The present work is one such determining step to further the development of latent heat storage systems based on optical charging route. Herein, we propose PCM seeded with thermochromic nanoparticles as potential novel nano-PCM with efficient and accelerated solar to latent heat storage capability.

Thermochromic nanoparticles have been shown to possess optical properties which are strong functions of temperature. Essentially, these exhibit two drastically different sets of optical properties in the neighbourhood of a unique temperature called "transition temperature ($T_{tr}$)". For instance, at temperatures below the transition temperature, these may possess a particular colour, and as the temperature is elevated to the transition temperature, these particles start to change their original colour and transforms to a different colour (just a few degrees above the transition temperature) [16, 17]. This concept has been recently put to use in the realms of cancer treatment (via photo-thermal therapy, to ensure deeper penetration of light for treatment of malignant tissues and decreasing the heating effect on nearby healthy tissues) [18-22] and green building design (particularly glazing for windows, to ensure thermal comfort to the building occupants) [23-25].

Building on the aforementioned ideas, the present work essentially investigates the candidature of thermochromic nanoparticles in efficient and accelerated latent heat storage of solar energy.

In particular, an attempt has been made to understand the intricately coupled heat and momentum transport mechanisms involved in optical charging of thermochromic nanoparticles seeded PCM. In addition to photon-nanoparticle interactions (and subsequent redistribution of energy through conduction and convection), thermochromism exhibited by the novel nano-PCM makes tracking of melting front more involved. Furthermore, to investigate the bounds of performance of the thermochromic nanoparticles seeded PCM; two unique situations have been critically investigated.

(a) Thermochromic nanoparticles assisted optical charging (TNP-OC) with gradual transition (GT) in optical properties at transition temperature; now on referred to as TNP-OC (GT). This essentially simulates thermochromism assisted photon transport (gradual transition).

(b) Thermochromic nanoparticles assisted optical charging (TNP-OC) with sudden transition (ST) in optical properties at transition temperature; now on referred to as i.e. TNP-OC (ST). This essentially simulates thermochromism assisted photon transport (sudden transition)

Although, the case of TNP-OC (GT) presents a more realistic practical situation; nevertheless, the TNP-OC (ST) case serves as a reference case and essentially represents the upper performance bound of TNP-OC route.

For a more holistic performance evaluation, the cases of thermal charging (referred to as TC) and nanoparticles assisted optical charging (photon transport without thermochromism assistance, referred to as NP-OC) have also been analysed. Moreover, a comprehensive comparative performance evaluation in terms of melting rate, latent heat storage capacity, spatial liquid fraction, and temperature distribution for a host of boundary conditions, solar incident flux values and nanoparticle volume fractions has been presented for all the aforementioned charging routes (viz., TNP-OC (GT), TNP-OC (ST), TC, and NP-OC).

## 2. BASIC CONCEPTS GOVERNING PCM CHARGING AND DISCHARGING ROUTES

### 2.1 Constructional details and conceptual design considerations

Figure 1 details conceptual designs of different latent heat storage systems studied in the present work. The basic philosophy underlying the studied configurations is to understand and draw qualitative as well as quantitative comparison among various possible routes that could be engineered for latent heat storage of the incident solar energy.

In particular, optical charging route (with and without thermochromism assistance) has been dealt with. Moreover, as pointed out earlier, the case of thermal charging (TC) could also be simulated by increasing the nanoparticles concentration to high values so that the photo-thermal energy conversion is limited to the first few layers of nano-PCM only. Therefore, in essence thermal charging and optical charging (with and without thermochromism assistance) cases have been critically and comprehensively analysed.

Looking into Figs. 1(d) - (f), clearly, the primary parameter that dictates the melting rate is the absorptivity in the solar irradiance region. In NP-OC charging, the greater the absorptivity (i.e., the absorption coefficient of the nanoparticles) of the nano-PCM, lesser is the depth up till which photon-nanoparticle interactions happen and vice a versa. Subsequent to this depth, the predominant mode of heat transfer is conduction (assisted by convection transport depending on the boundary conditions). On the other hand, in the case of thermochromism assisted photo-thermal energy conversion (TNP-OC), (for a given nanoparticles volume fraction, i.e., absorption coefficient) the depth up till which photon-nanoparticle interactions happen is much greater than the NP-OC case. For instance, in case of sudden transition case (TNP-OC (ST)), the absorptivity sharply drops down at the transition temperature; hence, making the way for photons to interact with the next layer of PCM. In case of gradual transition (TNP-OC (GT)), the absorptivity decreases linearly with temperature; allowing gradual increase in the photon-nanoparticle interactions. Thus, overall, thermochromism helps in careful control of the absorptivity and hence significantly accelerates the melting rate of the PCM without requiring the attainment of high temperatures.

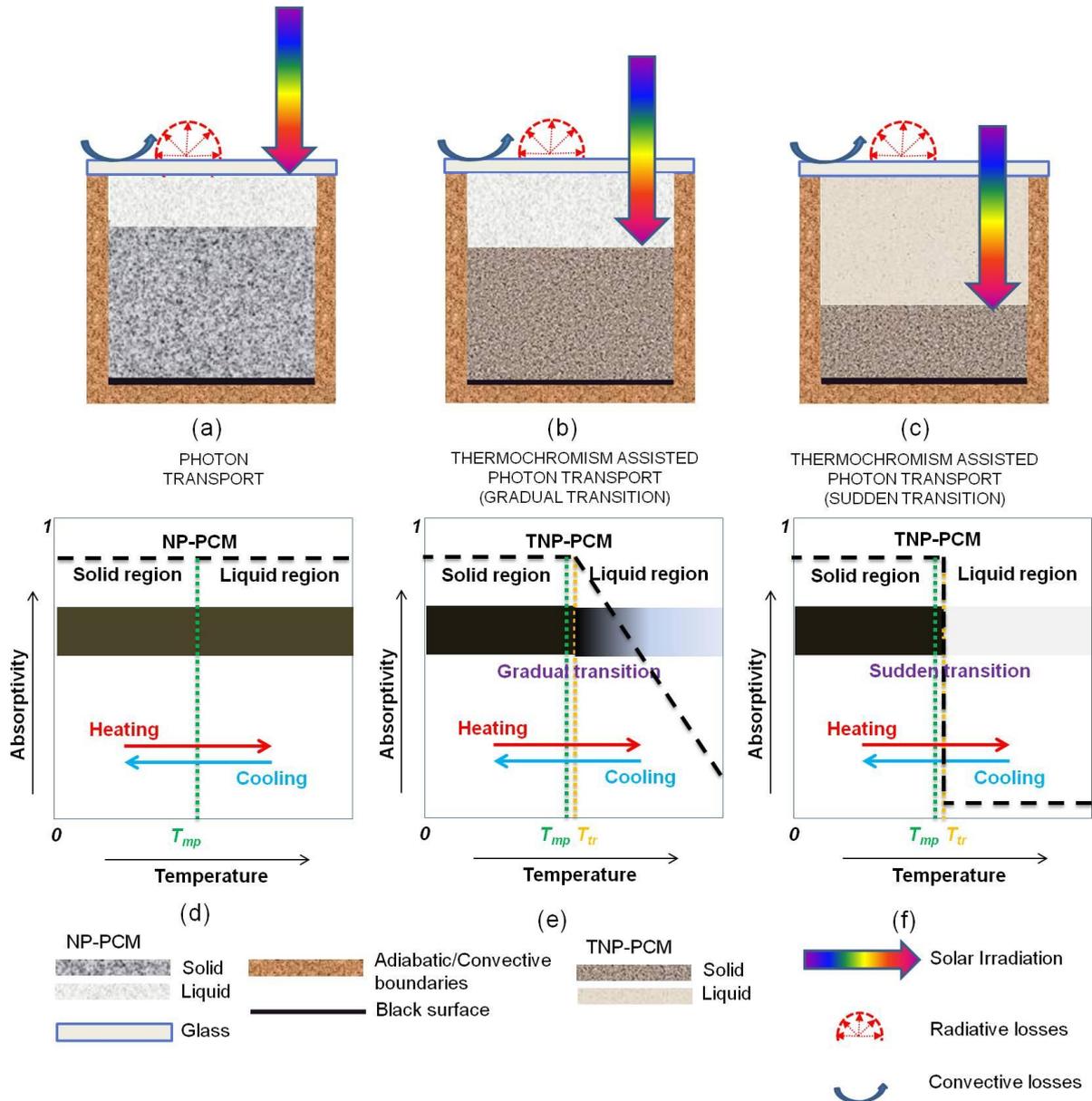

Fig. 1 Schematics showing solar irradiance interaction with (a) nanoparticles laden PCM (NP-PCM, simulating NP-OC and TC charging cases @ low and high nanoparticles concentrations respectively), (b) thermochromic nanoparticles laden PCM (TNP-PCM, simulating TNP-OC (GT) charging case, (c) thermochromic nanoparticles laden PCM (TNP-PCM, simulating TNP-OC (ST) charging case; schematics showing qualitative trends of absorptivity as a function of temperature for (d) NP-OC and TC charging cases, (e) TNP-OC (GT) charging case, and (f) TNP-OC (ST) charging case. Here, $T_{mp}$ and $T_{tr}$ represent the melting temperature of the PCM and the transition temperature of thermochromic nanoparticles respectively.

## 2.2 Identifying predominant heat transfer mechanisms dictating the melting of PCM

Figures 2(a), 2(b), and 2(c) show the predominant heat transfer mechanisms operational in thermal charging (TC), photon-transport based optical charging (NP-OC), and thermochromism assisted photon transport based optical charging (TNP-OC) routes respectively. It may be noted that at high values of nanoparticles volume fraction, the case of NP-OC approaches the case of thermal charging (TC). Clearly, in thermal charging route,

conduction is the predominant mode of heat transfer. On the other hand, the TNP-OC case ensures much deeper penetration of sunlight (and hence accelerated melting rate) owing to increased "photon-nanoparticle interactions" (hence, "radiative heat transfer" being the predominant mode of heat transfer mechanisms).

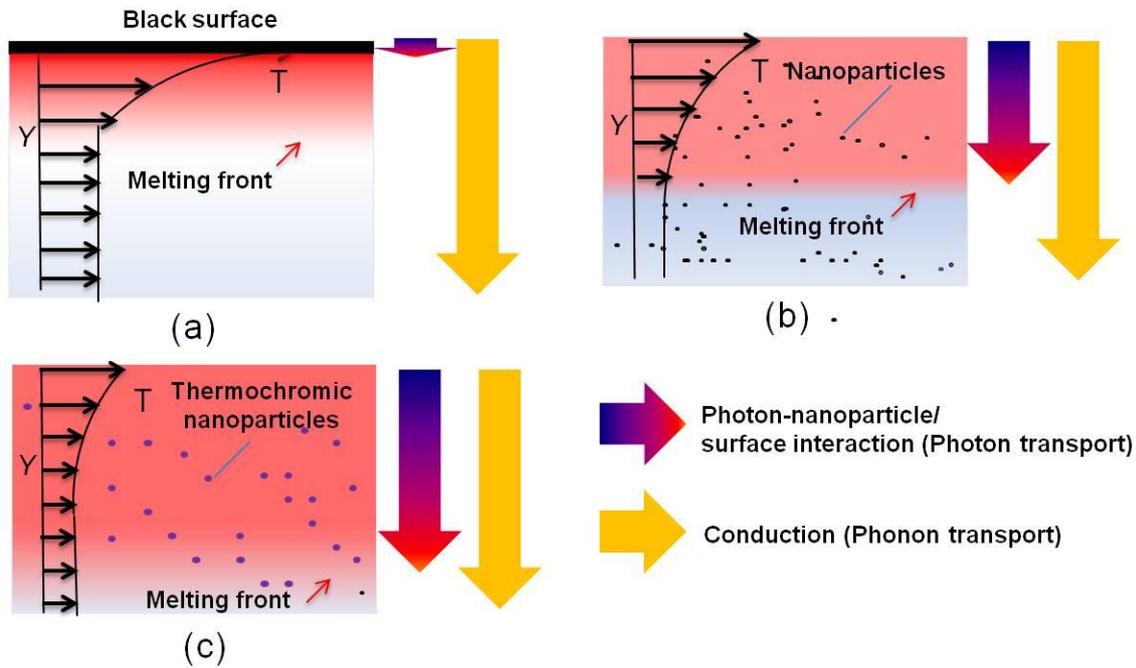

Fig. 2 Schematic diagrams showing the basic concepts of (a) thermal charging (TC), (b) photon-transport based optical charging (NP-OC), and (c) thermochromism assisted photon transport based optical charging (TNP-OC).

## 3. THEORETICAL MODELING FRAMEWORK
### 3.1 Underlying assumptions
- Incident solar energy flux is assumed to strike normally at the top surface of enclosure [26].
- Bottom surface of the enclosure is considered to be perfectly black; i.e., it absorbs all the solar radiation that is able to reach the bottom surface of enclosure.
- NP-PCM and TNP-PCM in liquid forms have been modelled as single phase semi-transparent participating media [26-30].
- Thermophysical properties of nano-PCMs, viz., NP-PCM and TNP-PCM have been assumed to be same as that of the pristine PCMs [26-30].

### 3.2 Theoretical modelling of momentum and heat transport mechanisms
Mass (Eq. (1)), momentum (Eqs. (2) and (3)) and energy (Eq. (6)) conservation equations have been solved to compute flow (in liquid phase) and temperature fields (both in liquid as well as solid phases) and to track the solid liquid interface.

$$\frac{\partial(\rho u)}{\partial x} + \frac{\partial(\rho v)}{\partial y} = 0, \tag{1}$$

$$\frac{\partial(\rho u)}{\partial t} + \frac{\partial(\rho u u)}{\partial x} + \frac{\partial(\rho v u)}{\partial y} = -\frac{\partial p}{\partial x} + \mu(\frac{\partial^2 u}{\partial x^2} + \frac{\partial^2 u}{\partial y^2}) + S_x, \tag{2}$$

$$\frac{\partial(\rho v)}{\partial t}+\frac{\partial(\rho uv)}{\partial x}+\frac{\partial(\rho vv)}{\partial y}=-\frac{\partial p}{\partial y}+\mu(\frac{\partial^2 v}{\partial x^2}+\frac{\partial^2 v}{\partial y^2})+S_y, \qquad (3)$$

where $u$ and $v$ are the velocity components in $x$ and $y$ directions; $\mu$ is the dynamic viscosity; $p_{nf}$ is the effective pressure; $S_x$ and $S_y$ represent the Boussinesq source terms in $x$ and $y$ directions respectively and are mathematically given by Eqs. (4) and (5) as

$$S_x = Au \qquad (4)$$

$$S_y = \rho_{ref}\beta g(T-T_{ref}) \qquad (5)$$

where $A(=-C(1-\varepsilon))$ is the porosity function and has been employed to ensure zero velocities in solid region; $C$ is the morphology constant (= 1.60 x 10$^6$); $\varepsilon$ (=$\Delta H/LH$) is the liquid fraction; $\beta$ is the coefficient of thermal expansion, $\rho_{ref}$ is the reference density of, $g$ is the acceleration due to gravity, $T$ is the local temperature and $T_{ref}$ is the reference temperature.

$$\frac{\partial(\rho H)}{\partial t}+\frac{\partial(\rho uH)}{\partial x}+\frac{\partial(\rho vH)}{\partial y}=k(\frac{\partial^2 H}{\partial x^2}+\frac{\partial^2 H}{\partial y^2})+S^H+S^T \qquad (6)$$

where $S^H$ and $S^T$ are the source terms representing latent heat absorption and radiative flux decay respectively within the physical domain of enclosures and expressed as Eqs. (7) and (8).

$$S^H = \rho\frac{\partial \Delta H}{\partial t} \qquad (7)$$

$$S^T = -\frac{\partial q_r}{\partial y} \qquad (8)$$

where $\Delta H$ represents the enthalpy for each control volume and $q_r$ is the radiative flux. Herein, absorption coefficient values employed for calculation of radiative flux for NP-OC, TNP-OC (GT), and TNP-OC (ST) charging cases are given by Eqs. (9), (10), and (11) respectively.

$$K_{abs,T} = K_{abs}|_{nano-PCM} \quad \text{for any } T \qquad (9)$$

$$K_{abs,T} = K_{abs,T_{tr}} - (T-T_{tr}) \qquad (10)$$

$$K_{abs,T} = K_{abs}|_{nano-PCM} \quad \text{for } T < T_{tr} \qquad (11a)$$

$$K_{abs,T} = K_{abs}|_{pristine-PCM,liquid} \quad \text{for } T \geq T_{tr} \qquad (11b)$$

It may be further noted that in the case of thermal charging, the absorption coefficient takes a constant value of 10000m$^{-1}$.

The aforementioned governing equations are subjected to the following boundary conditions:

$$u = v = 0, \text{ for } x = 0 \text{ and } x = B \quad (12a)$$

$$u = v = 0, \text{ for } y = 0 \text{ and } y = D \quad (12b)$$

$$T(y,0) = T_i = T_{amb} \quad (12c)$$

At the top irradiated surface, convective and radiative losses are calculated using Eq. (13a)

$$-k_T a_T \frac{\partial T}{\partial y}\bigg|_{y=0} = h_T a_T (T_{sf} - T_{amb}) + \sigma a_T (T_{sf}^4 - T_{amb}^4) = Q''_{loss\_conv} + Q''_{loss\_rad} \quad (13a)$$

where, $h_{t\_conv} = 0.27 Ra^{1/4} \ldots\ldots 10^5 \leq Ra \leq 10^{10}, \Pr = 0.71$ [31] $\quad (13b)$

$$h_{t\_rad} = \varepsilon_g \sigma (T_{sf}^2 + T_{amb}^2)(T_{sf} + T_{amb}) \quad (13c)$$

At the bottom, left, and right boundaries two cases have been considered

$$-k_B a_B \frac{\partial T}{\partial y}\bigg|_{y=D} = h_B a_B (T_{sf} - T_{amb}) \quad (14a)$$

$$-k_L a_L \frac{\partial T}{\partial y}\bigg|_{x=0} = h_L a_L (T_{sf} - T_{amb}) \quad (14b)$$

$$-k_R a_R \frac{\partial T}{\partial y}\bigg|_{x=B} = h_R a_R (T_{sf} - T_{amb}) \quad (14c)$$

Case 1: Adiabatic boundaries
Here, convective heat transfer coefficients corresponding to left, right and top walls of enclosure are $h_L = h_R = h_T = 0$ Wm$^{-2}$K$^{-1}$. This case essentially represents the "charging" of the thermal storage system; i.e., solar energy is being converted and stored in the form of latent heat of PCM.

Case 2: Convective boundaries
Here, convective heat transfer coefficients corresponding to left, right and top walls of enclosure are

a) $\quad h_L = h_R = h_B = 100$ Wm$^{-2}$K$^{-1}$ $\quad (15a)$

b) $\quad h_L = h_R = h_B = 1000$ Wm$^{-2}$K$^{-1}$ $\quad (15b)$

This case represents "simultaneous charging and discharging" of the thermal storage system. The values of heat transfer coefficient have been taken in a range to simulate discharging at the walls by free and forced movement of liquids [32].

Mathematically, the latent heat storage capacity (LHSC) and sensible heat discharging capacity (SHDC) are given by Eqs. (16) and (17) respectively.

$$LHSC(\%) = \frac{\sum_{i=1,j=1}^{Nx,Ny} \varepsilon_{ij} \Delta H_{ij} \rho dxdy}{\sum_{i=1,j=1}^{Nx,Ny} \Delta H_{ij} \rho dxdy} \times 100 \quad (16)$$

$$SHDC(\%) = 1 - \left[ \frac{(Q^{"}_{loss\_conv} + Q^{"}_{loss\_rad})_T}{\sum_{all\_walls} -ka_s \frac{\partial T}{\partial j}} \right] \times 100 \quad (17)$$

where $j = x$ and $y$ for left and right, and bottom and top walls respectively.

Furthermore, to aid comparison of optical charging routes with the thermal charging route, we define $\delta Y_{TC}$ (%), $\delta SHDC_{TC}$ (%) and $\delta LHSC_{TC}$ (%) as follows:

$$\delta Y_{TC}(\%) = \frac{Y_{OC} - Y_{TC}}{Y_{TC}} \times 100 \quad (18a)$$

$$\delta LHSC_{TC}(\%) = \frac{LHSC_{OC} - LHSC_{TC}}{LHSC_{TC}} \times 100 \quad (18b)$$

$$\delta SHDC_{TC}(\%) = \frac{SHDC_{OC} - SHDC_{TC}}{SHDC_{TC}} \times 100 \quad (18c)$$

These represent the enhancements in melting front movement, sensible heat discharging capacity and latent heat storage capacity relative to the thermal charging route.

### 3.3 Numerical modelling

Governing differential equations for mass, momentum, and energy are numerically solved using finite volume approach. "Semi-implicit method for pressure-linked equations" (SIMPLE) and "enthalpy porosity method" have been employed to numerically solve momentum and energy equations respectively.

Figure 3 presents the computational algorithm (along with the typical values of properties/parameters) implemented in MATLAB® to compute the temperature and flow fields, phase front, liquid fraction and latent heat storage capacity. In particular, first-order upwind scheme has been applied to the convective terms of the momentum and energy equations whereas central difference scheme has been employed for diffusion terms of the equations [33-35].

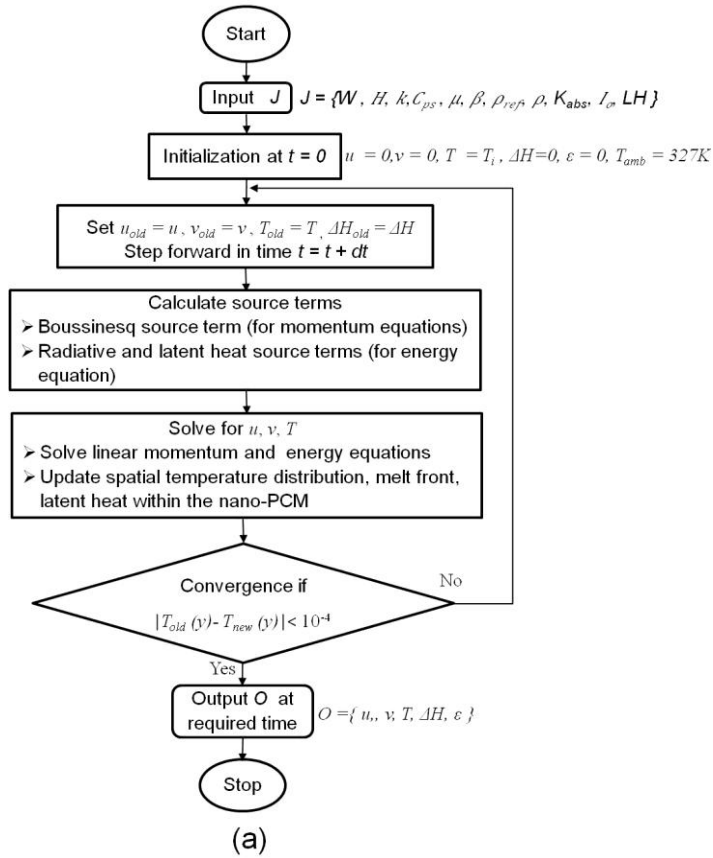

Fig. 3 (a) Flow chart showing the algorithm implemented in MATLAB to compute temperature and flow fields, phase front, liquid fraction, sensible heat discharging capacity, and latent heat storage capacity, and (b) typical values of properties/parameters employed.

## 3.4 Grid independence test

To ensure that the results are independent of grid size, a grid independence test has been carried out for square enclosure with dimensions ($D = 0.04$ m, $B = 0.04$ m) corresponding to surface absorption mode. Herein, mid plane temperatures within the enclosure at $I_o = 1000$ Wm$^{-2}$, and $K_{abs} = 100000$ m$^{-1}$ are reported for grid sizes $20 \times 20$, $40 \times 40$ and $60 \times 60$ (see Fig. 4). The results corresponding to grid sizes $40 \times 40$ and $60 \times 60$ seem to be closer, therefore, $40 \times 40$ grid size has been chosen for the present work.

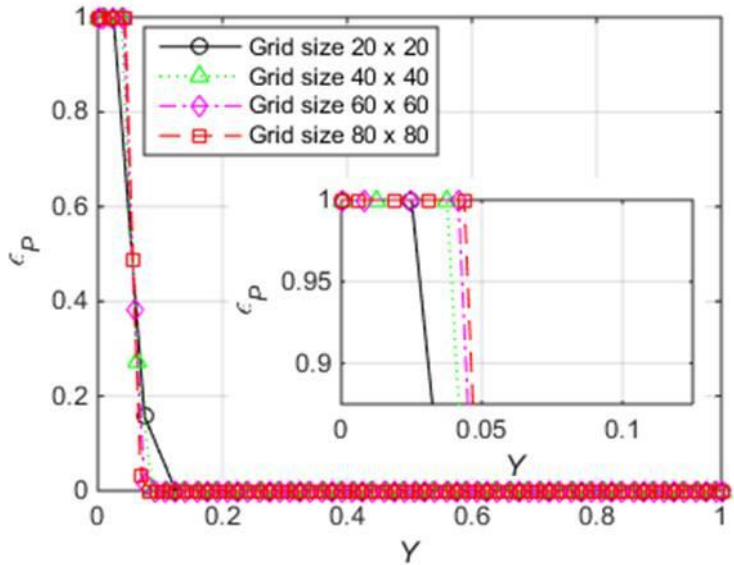

Fig. 4 Mid-plane liquid fraction along receiver depth for various grid sizes (20 × 20, 40 × 40, and 60 × 60).

### 3.5 Validation of Mathematical Model

The current developed model using enthalpy porosity technique has been validated with the experimental [36] and numerical [33] results for melting of Gallium. Herein, differential side heating of Gallium was studied in rectangular enclosure ($B$ = 8.89 cm, $D$ = 6.35 cm) having adiabatic top and bottom walls, with left as hot and right as cold wall. Results have been reported for "melt front" at different instants in time (i.e., 2, 6, 10 and 19 minutes) as shown in Fig. 5.

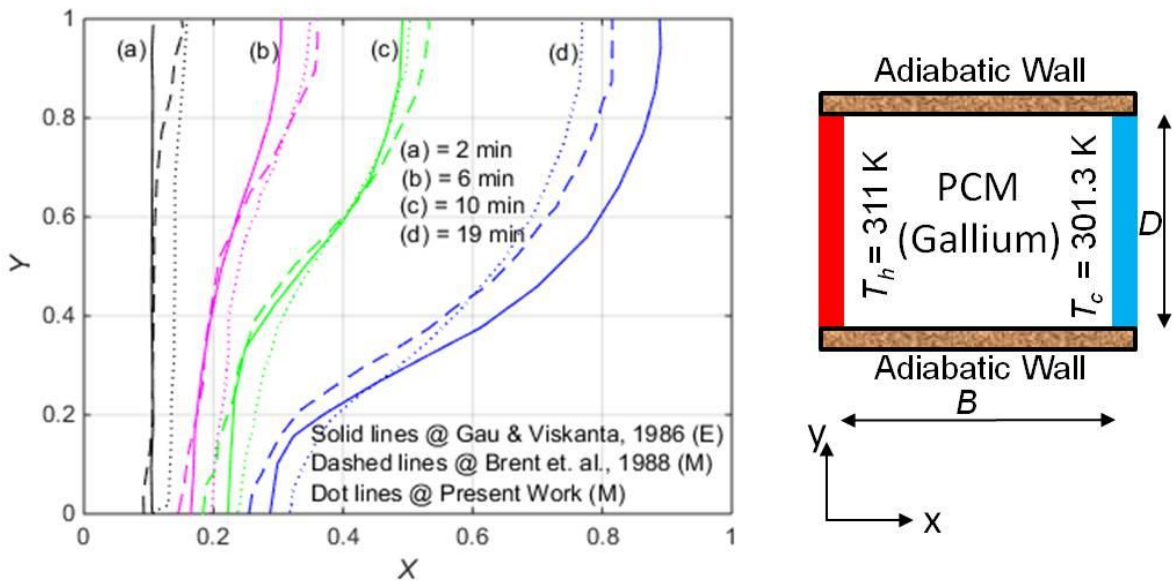

Fig. 5 Comparison of "melt front" values (for present model) with the experimental and theoretical values of Gau & Viskanta, 1986 [36] and Brent et al., 1988 [33] respectively.

## 4. RESULTS AND DISCUSSION

**4.1 Assessing the fundamental limits of latent heat storage for thermal and optical charging of routes (TC, NP-OC, TNP-OC (GT), and TNP-OC (ST)) under adiabatic boundary conditions ($h_L = h_R = h_T = 0$ Wm$^{-2}$K$^{-1}$).**

*4.1.1 Effect of charging route on the intensity attenuation in nano-PCM along the depth direction*

Figure 6 (a) - (d) show the non-dimensional flux values along the depth direction for the optical charging route (NP-OC) for different values of nano-PCM absorption coefficient. Clearly, penetration depth of the incident solar flux decreases with increase in the value of absorption coefficient, and approaches the case of surface absorption at very high values of absorption coefficient ($K_{abs}$ = 10000m$^{-1}$). On the other hand, in the case of themochromism assisted optical charging (TNP-OC), the penetration depth is a function of absorption coefficient as well as temperature (represented by curves at different instants in time, see Fig. 6(e) - 6(l)). Particularly, in the case of TNP-OC (ST), a unique feature can be seen in relation to the penetration depth. Herein, irrespective of the value of the absorption coefficient (except for the extreme value of absorption coefficient, i.e., $K_{abs}$ = 10000m$^{-1}$), the solar flux is able to make its way all through up till the bottom (i.e., *R* has a finite large value @ *Y* = 0, see Fig. 6(i) - (k)). Thus, paving the way for the "secondary melt front" to originate at the bottom. It may further be noted that even for $K_{abs}$ = 10000m$^{-1}$, the secondary melt front shall be observed as we further march in time. Figure 7 shows the primary and secondary melt fronts for a representative case of TNP-OC (ST) [wherein the solar flux (= 10000Wm$^{-2}$)] for different values of nanoparticles absorption coefficient.

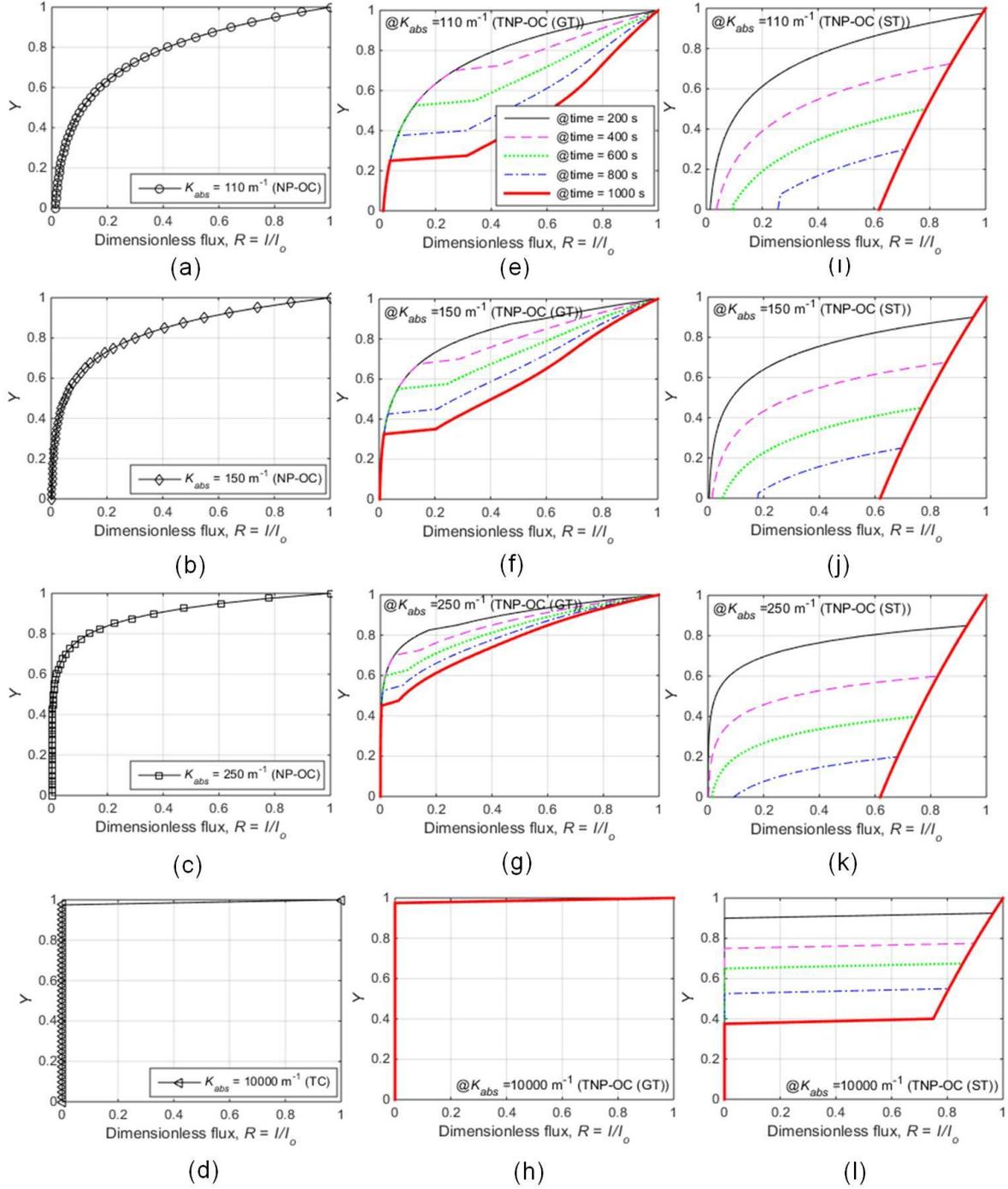

Fig. 6 Intensity attenuation for different absorption coefficients approximating (a) NP-OC, $K_{abs,}$ =110 m$^{-1}$ (b) NP-OC, $K_{abs,}$ =150 m$^{-1}$ (c) NP-OC, $K_{abs,}$ =250 m$^{-1}$ (d) NP-OC, $K_{abs,}$ = 10000 m$^{-1}$ (e) TNP-OC (GT), $K_{abs,}$ =110 m$^{-1}$ (f) TNP-OC (GT), $K_{abs,}$ =150 m$^{-1}$ (g) TNP-OC (GT), $K_{abs,}$ =250 m$^{-1}$ (h) TNP-OC (ST), $K_{abs,}$ =10000 m$^{-1}$ (i) TNP-OC (ST), $K_{abs,}$ =110 m$^{-1}$ (j) TNP-OC (ST), $K_{abs,}$ =150 m$^{-1}$ (k) TNP-OC (ST), $K_{abs,}$ =250 m$^{-1}$ (l) TNP-OC (ST), $K_{abs,}$ =10000 m$^{-1}$.

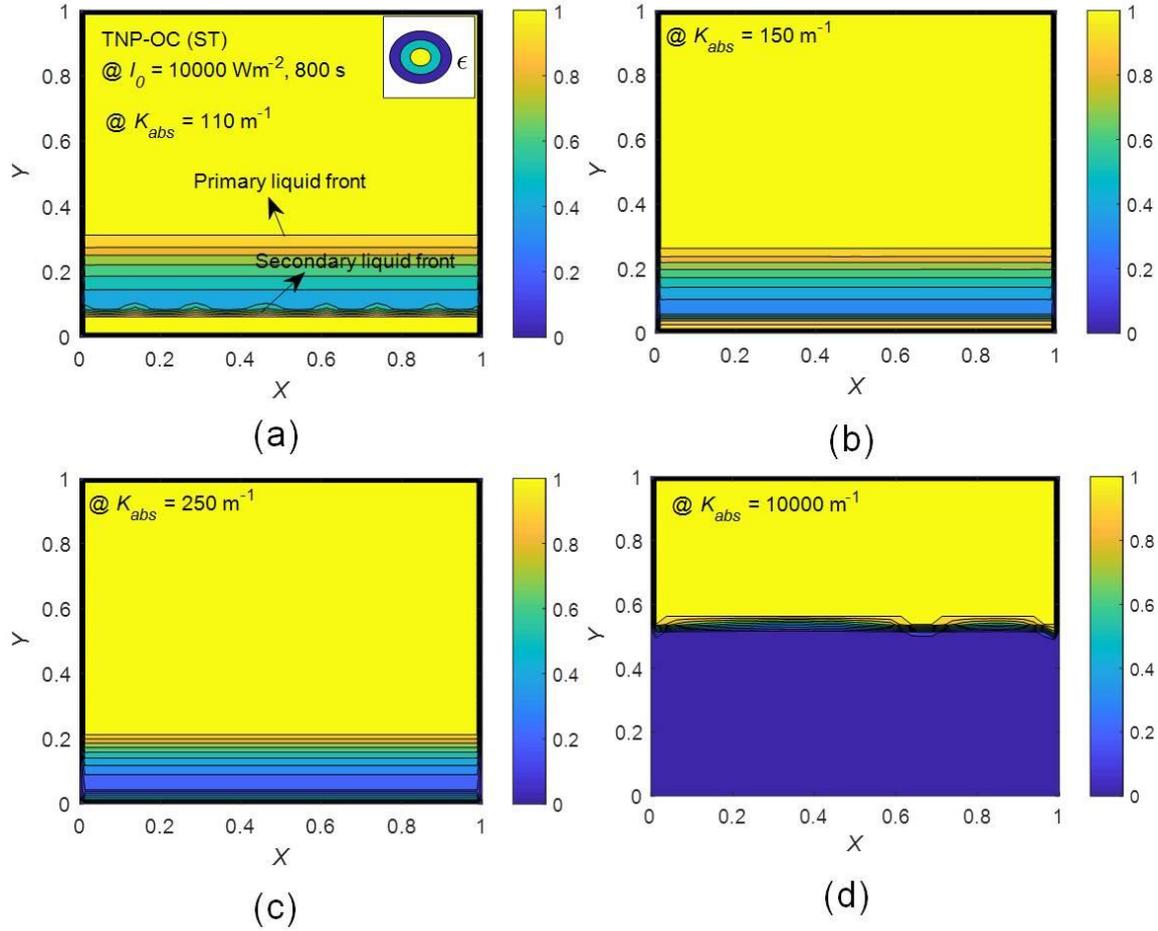

Fig. 7 Contours of liquid fraction ($\varepsilon$) showing primary and secondary liquid fronts in the case of TNP-OC (ST), corresponding to $I_o = 10000$ Wm$^{-2}$, $t = 800$ s under adiabatic boundary conditions ($h_L = h_R = h_T = 0$ Wm$^{-2}$K$^{-1}$) for (a) $K_{abs} = 110$ m$^{-1}$ (b) $K_{abs} = 150$ m$^{-1}$ (c) $K_{abs} = 250$ m$^{-1}$, and (d) $K_{abs} = 10000$ m$^{-1}$.

*4.1.2 Temporal evolution of primary melting (or liquid) front and latent heat storage capacity (LHSC)*

Figure 8 compares temporal evolution of primary liquid front for thermal and optical charging routes with incident flux, $I_o = 10000$ Wm$^{-2}$. For initial time period (of approximately 200 s), thermal charging route offers fast melting relative to the optical charging route. This may be attributed to the enhanced absorption (and that too confined to only a few layers of the nano-PCM) owing to very large value of absorption coefficient ($K_{abs} = 10000$ m$^{-1}$). In thermal charging case, although, the top layer temperature increases very rapidly (see Fig. 9), this does not percolate down the subsequent layer along the depth direction due to inherently low thermal conductivity of the nano-PCM (see Fig.10). On the other hand, the optical charging routes take time to get started, however, after some time has elapsed, accelerated melting rate is observed and these overtake the thermal charging route. Similar trends (but with less pronounced magnitudes) are observed for lower incident flux values as well (see Fig. A1). Furthermore, for completeness, the mid-plane liquid fraction distribution along the depth direction has been plotted in Fig. A2.

Figure 11 compares the enhancements (relative to thermal charging) in terms of progression of the liquid front observed in various cases of optical charging for different values of absorption coefficient. Maximum improvements are observed in the case of TNP-OC (ST) followed by TNP-OC (GT) and NP-OC; also the magnitudes of enhancements increase with increase in the incident flux values. Interestingly, in case of TNP-OC (ST), there exits an optimum value of nanoparticles absorption coefficient at which the enhancement is maximum ($K_{abs}$ = 250m$^{-1}$). This may be understood from the fact that, initially, with increase in absorption coefficient values, absorptivity and hence the temperature increases, owing to which sudden transition in optical properties happens early - thus exposing the subsequent layers to the incident flux and accelerating the melting process. However, on further increasing the absorption coefficient, the thermal losses increase significantly and thus results in lowering the enhancements. Therefore, there is a need to strike a balance between increasing absorptivity and thermal losses with increase in absorption coefficient values.

On the other hand, in the case of other optical charging routes (viz., TNP-OC (GT) and NP-OC), the magnitude of enhancement continuously decreases with increase in absorption coefficient values owing to increase in thermal losses with increase in absorption coefficient values.

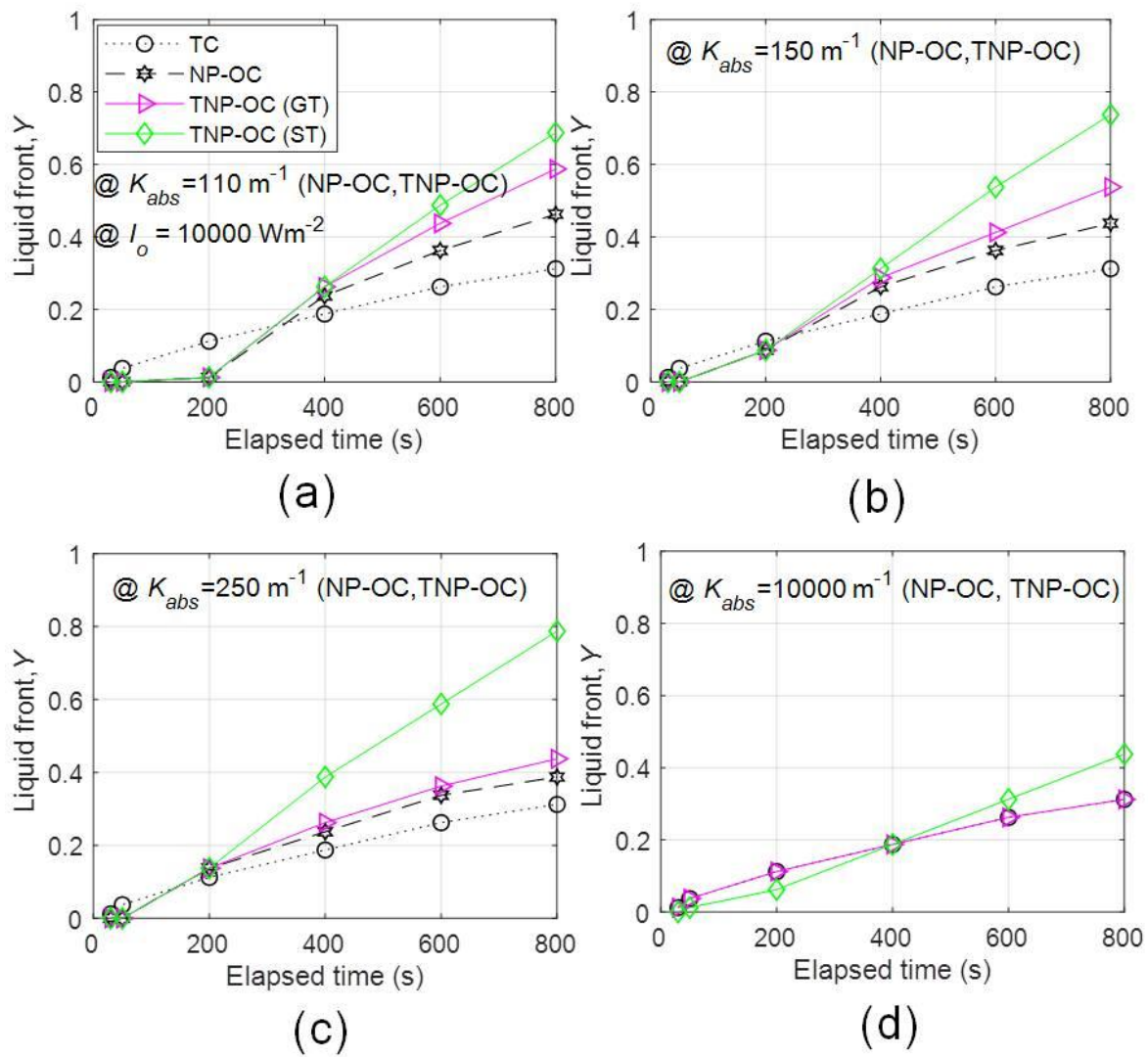

Fig.8 Comparison of midplane primary liquid front position for nano-PCM, TC ($K_{abs}$ = 10000 m$^{-1}$) and OC ( viz. NP-OC, TNP-OC (GT), TNP-OC (ST)) corresponding to flux $I_o$ = 10000 Wm$^{-2}$ (a) $K_{abs}$ = 110 m$^{-1}$ (b) $K_{abs}$ = 150 m$^{-1}$ (c) $K_{abs}$ = 250 m$^{-1}$ (d) $K_{abs}$ = 10000 m$^{-1}$.

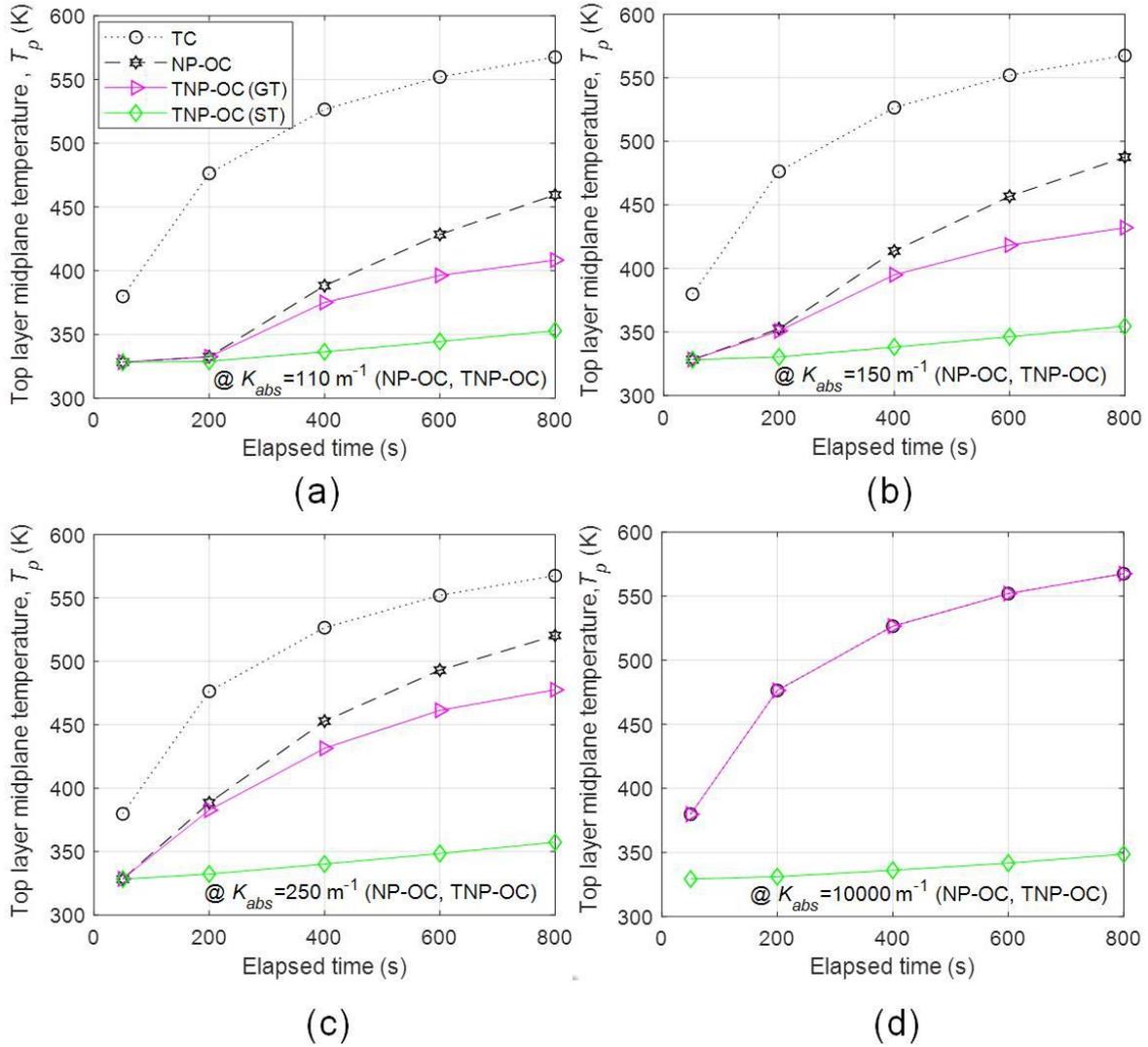

Fig. 9 Comparison of top layer mid-plane temperature for nano-PCM, TC ($K_{abs}$ = 10000 m$^{-1}$) and OC ( viz. NP-OC, TNP-OC (GT), TNP-OC (ST)) corresponding to flux $I_o$ = 10000 Wm$^{-2}$ (a) $K_{abs}$ = 110 m$^{-1}$ (b) $K_{abs}$ = 150 m$^{-1}$ (c) $K_{abs}$ = 250 m$^{-1}$ (d) $K_{abs}$ = 10000 m$^{-1}$.

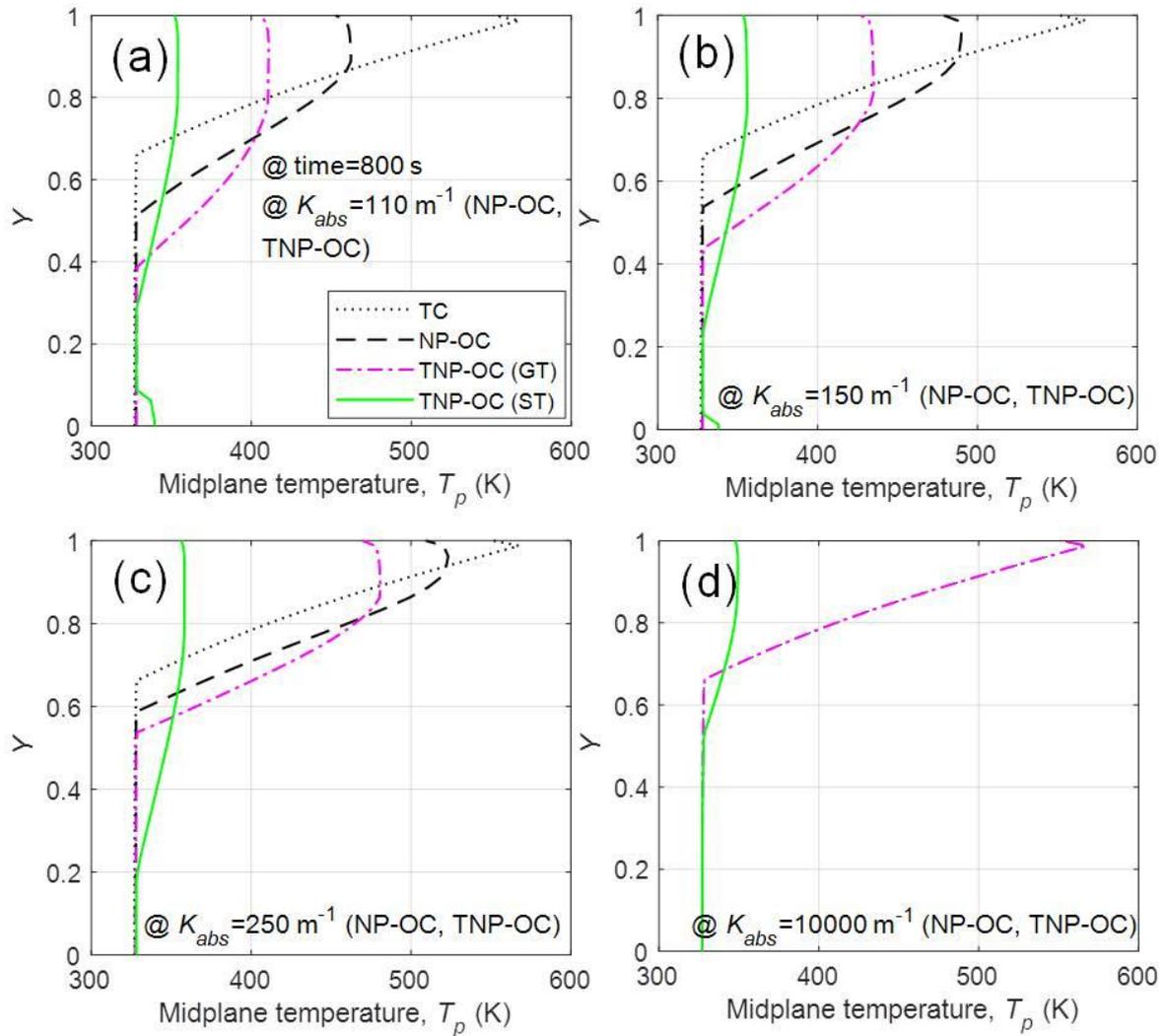

Fig. 10 Comparison of midplane temperature distribution for nano-PCM, TC ($K_{abs}$ = 10000 m$^{-1}$) and OC ( viz. NP-OC, TNP-OC (GT), TNP-OC (ST)) corresponding to flux $I_o$ = 10000 Wm$^{-2}$ (a) $K_{abs}$ = 110 m$^{-1}$ (b) $K_{abs}$ = 150 m$^{-1}$ (c) $K_{abs}$ = 250 m$^{-1}$ (d) $K_{abs}$ = 10000 m$^{-1}$.

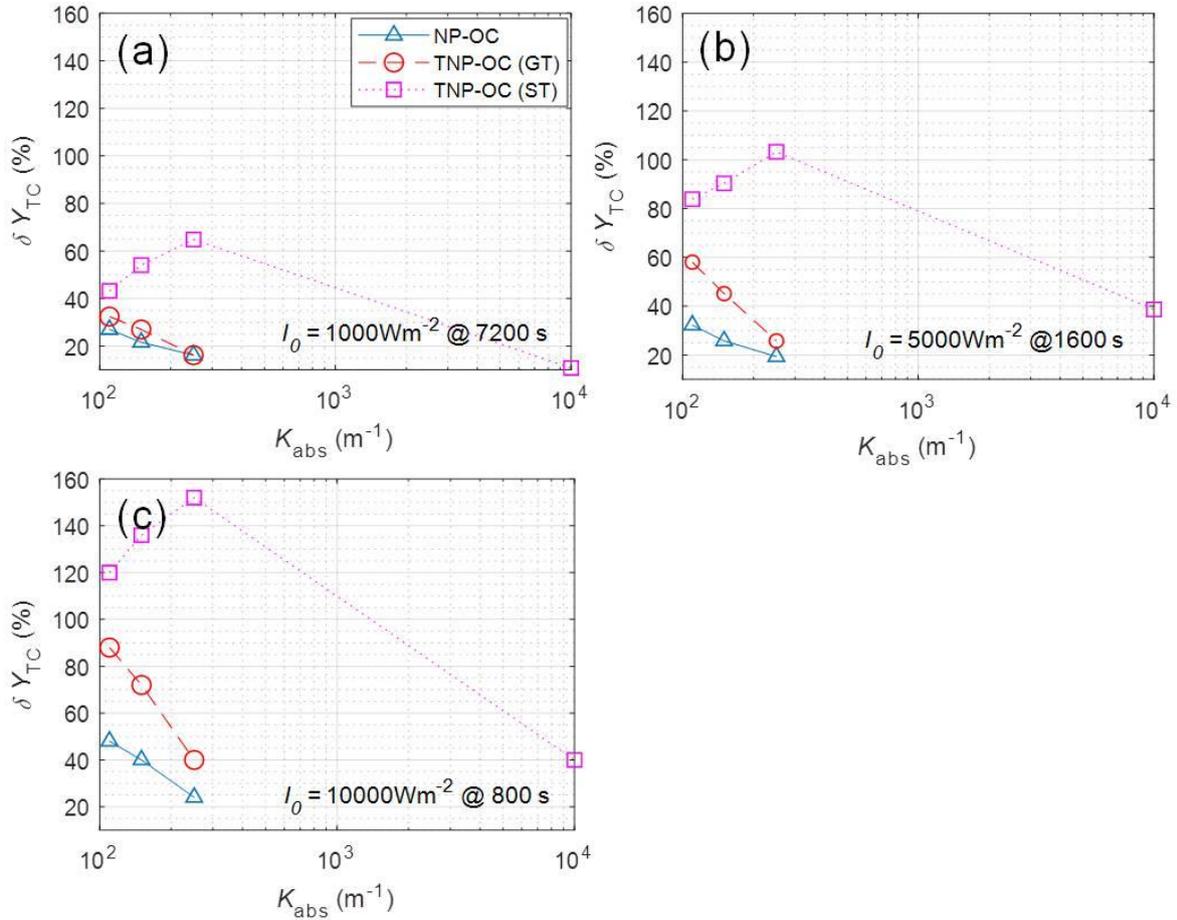

Fig. 11 Enhancements (relative to thermal charging) in terms of progression of the liquid front observed in various cases of optical charging for different values of absorption coefficient for (a) $I_0$ = 10000Wm$^{-2}$, (b) $I_0$ = 5000Wm$^{-2}$, and (c) $I_0$ = 1000Wm$^{-2}$

Looking into the temporal evolution of latent heat storage capacity (see Fig. 12) for a representative case of $I_0$ = 10000Wm$^{-2}$. For initial phase of melting (except for very high values of absorption coefficient ($K_{abs}$ = 10000 m$^{-1}$)), invariably, TNP-OC (ST) route offers the greatest advantage followed by TNP-OC (GT), NP-OC and TC routes of charging. Subsequently, even for $K_{abs}$ = 10000 m$^{-1}$, the aforementioned trend is followed. Figure 13 shows the percentage enhancements (in LHSC %) relative to the thermal charging (TC) case. The magniude of enhancement decreases with increase in absorption coefficient values. This may be be attributed to the increase in thermal losses with increase in absorption coefficient values.

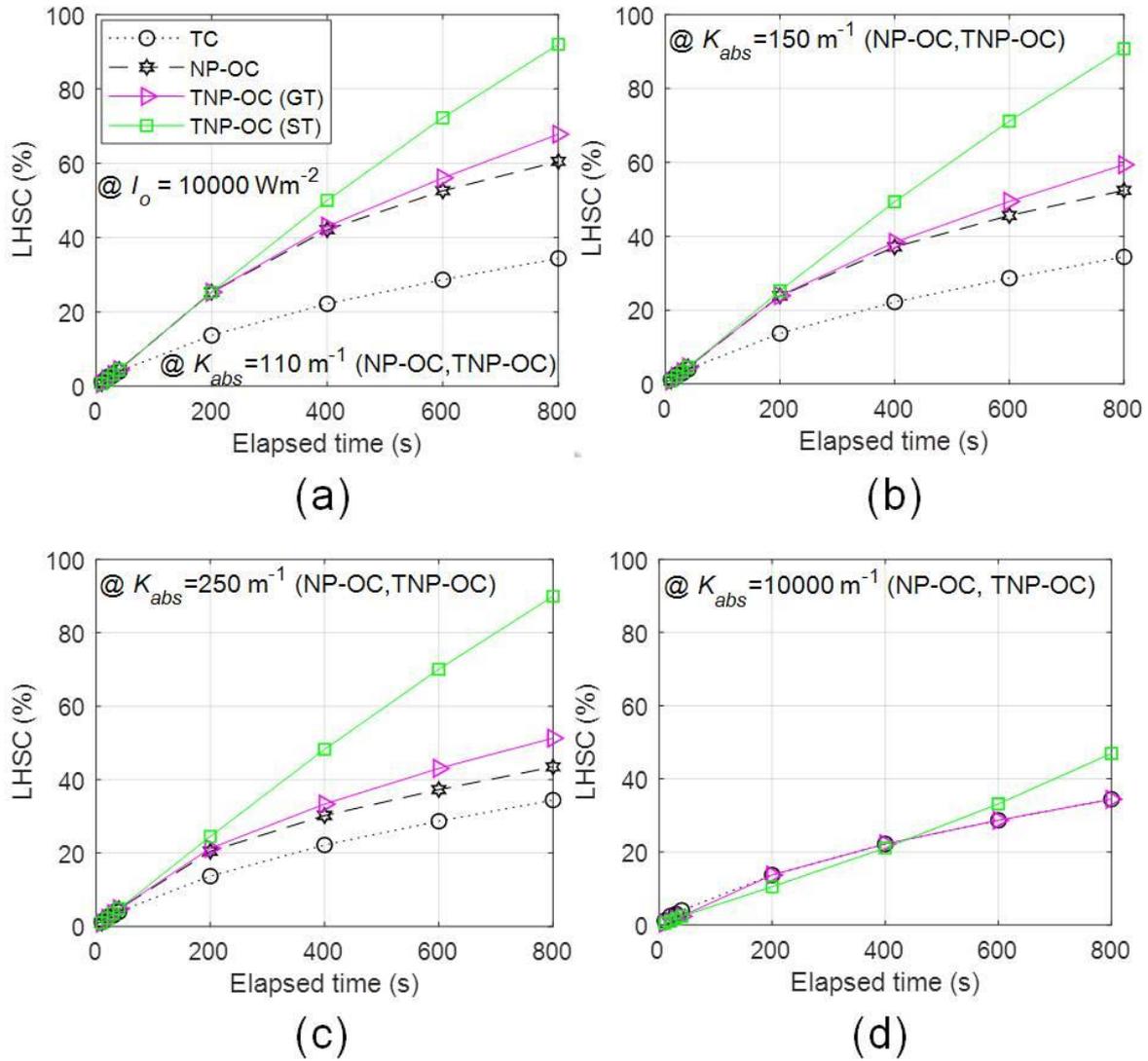

Fig. 12 Comparison of latent heat storage capacity for nano-PCM, TC ($K_{abs}$ = 10000 m$^{-1}$) and OC ( viz. NP-OC, TNP-OC (GT), TNP-OC (ST)) corresponding to flux $I_o$ = 10000 Wm$^{-2}$:
(a) $K_{abs}$ = 110 m$^{-1}$, (b) $K_{abs}$ = 150 m$^{-1}$, (c) $K_{abs}$ = 250 m$^{-1}$, and (d) $K_{abs}$ = 10000 m$^{-1}$.

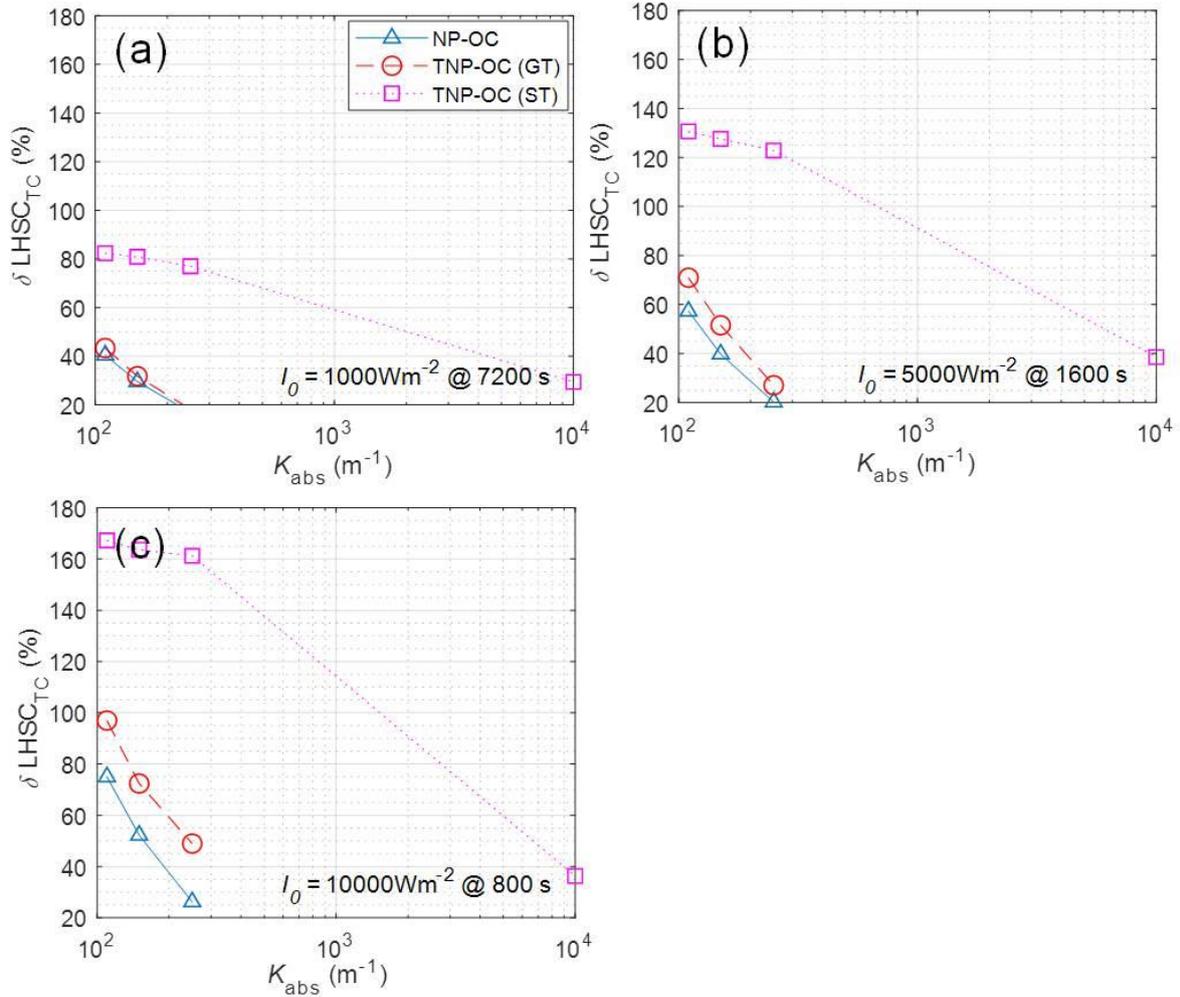

Fig. 13 Enhancements (relative to thermal charging) in terms of latent heat storage capacity (LHSC %) observed in various cases of optical charging for different values of absorption coefficient for (a) $I_0$ = 10000Wm$^{-2}$, (b) $I_0$ = 5000Wm$^{-2}$, and (c) $I_0$ = 1000Wm$^{-2}$

**4.2 Assessing the fundamental limits of latent and sensible heat storage for thermal and optical charging of routes (TC, NP-OC, TNP-OC (GT), and TNP-OC (ST)) under convective boundary conditions ($h_L \neq h_R \neq h_T \neq 0$ Wm$^{-2}$K$^{-1}$).**

Now that we have deciphered the latent heat storage capacity (LHSC %) under adiabatic boundary conditions; next, we quantitatively compare the simultaneous LHSC and SHDC, which simulates "simultaneous charging and discharging" of the latent heat storage system (see Fig. 14). Except for the TNP-OC (ST), in all other charging routes the LHSC (%) monotonically decreases with increasing absorption coefficient (owing to higher thermal losses at high absorption coefficient values) irrespective of the incident flux values. Whereas in case of TNP-OC (ST), at low incident flux values (i.e., 1000Wm$^{-2}$), initially, an increase in absorption coefficient value enhances the absorption capability and hence the temperature and LHSC (%). However, after a certain value of absorption coefficient, the thermal losses also tend to escalate - thus decreasing the LHSC (%) for high values of absorption coefficient. On the other hand, at higher incident flux values, TNP-OC (ST) follows similar trend i.e., LHSC (%) monotonically decreases with increasing absorption coefficient values.

Looking into extracting the sensible heat from the storage system, Fig. 15 compares SHDC (%) as a function of the absorption coefficient for different charging routes. Here, clearly, in all cases SHDC (%) monotonically decreases with increasing absorption coefficient owing to increased thermal losses at high values of absorption coefficient.

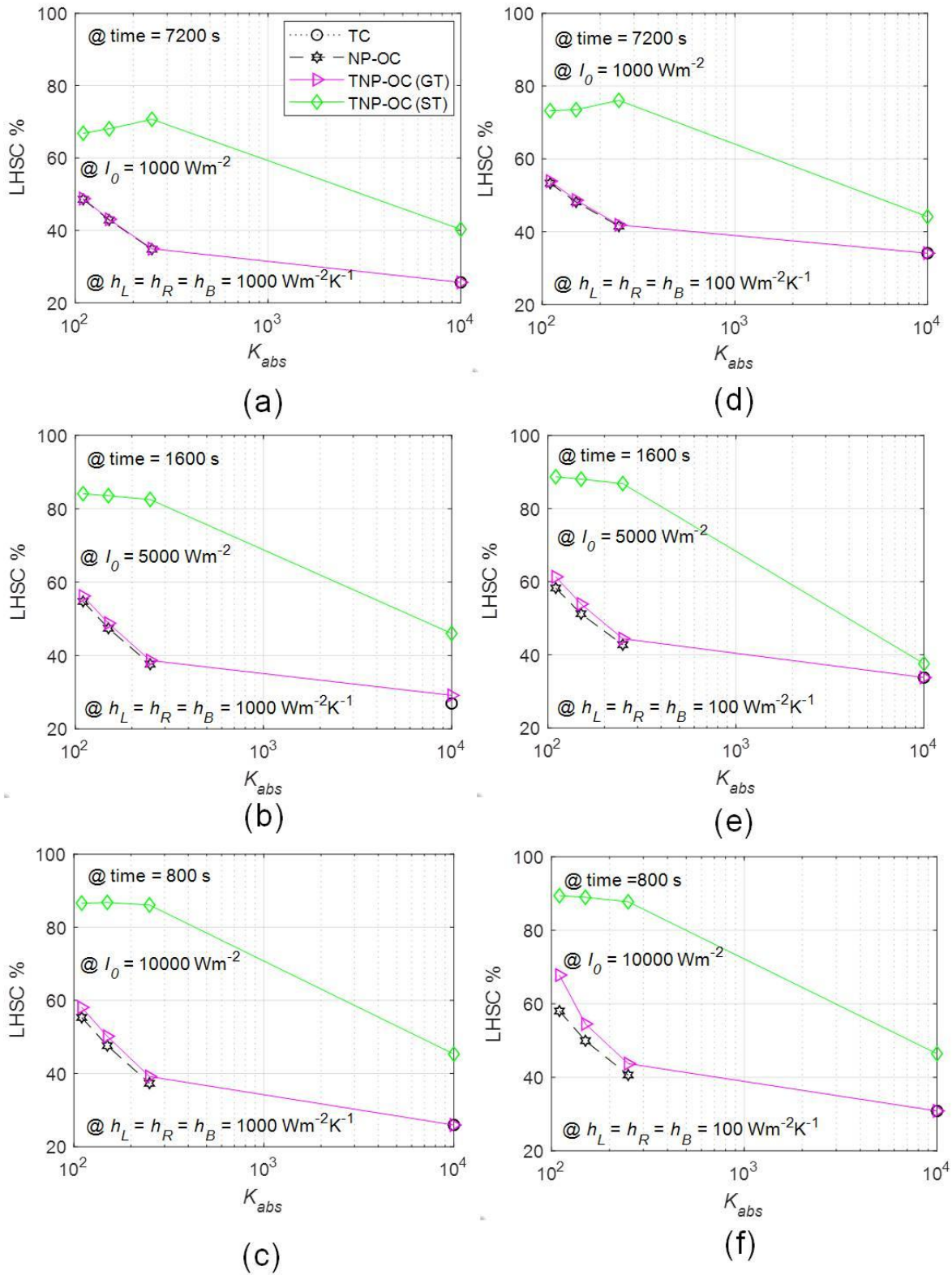

Fig. 14 Variation of LHSC (%) with absorption coefficients corresponding to convective boundary; $h_L = h_R = h_B = 100$ Wm$^{-2}$K$^{-1}$ @incident flux [(a) $I_o = 10000$ Wm$^{-2}$; (b) $I_o = 5000$ Wm$^{-2}$; (c) $I_o = 1000$ Wm$^{-2}$] and $h_L = h_R = h_B = 1000$ Wm$^{-2}$K$^{-1}$ @incident flux [(d) $I_o = 10000$ Wm$^{-2}$; (e) $I_o = 5000$ Wm$^{-2}$; (f) $I_o = 1000$ Wm$^{-2}$]

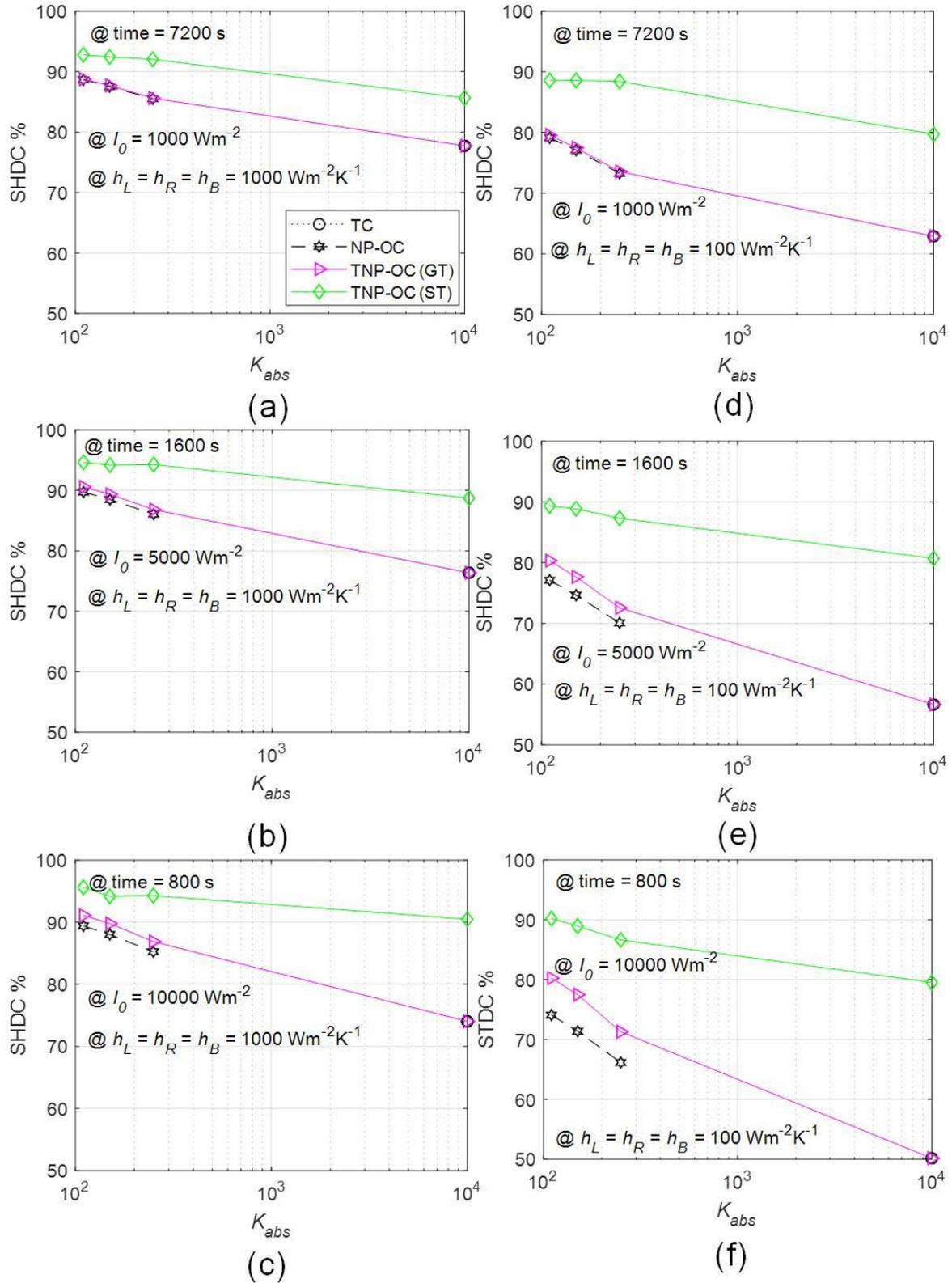

Fig. 15 Variation of SHDC (%) with absorption coefficients corresponding to convective boundary; $h_L = h_R = h_B = 100$ Wm$^{-2}$K$^{-1}$ @incident flux [(a) $I_o = 10000$ Wm$^{-2}$; (b) $I_o = 5000$ Wm$^{-2}$; (c) $I_o = 1000$ Wm$^{-2}$] and $h_L = h_R = h_B = 1000$ Wm$^{-2}$K$^{-1}$ @incident flux [(d) $I_o = 10000$ Wm$^{-2}$; (e) $I_o = 5000$ Wm$^{-2}$; (f) $I_o = 1000$ Wm$^{-2}$].

Relative to thermal charging (TC), thermochromism assited optical charging route, particularly TNP-OC (ST) case (@ $I_0$ =10000 Wm$^{-2}$) enhances the LHSC (%) and SHSC (%) by approximately by 221 % and 80 % respctively.

**CONCLUSIONS**

In the current work, a detailed opto-thermal modelling framework has been developed to understand and quantify coupled momentum and energy transport mechanisms governing various thermal (TC) and optical (NP-OC, TNP-OC (GT), and TNP-OC (ST)) charging routes. Critical investigation reveals that through thermochromism assisted photon transport significant improvements in the performance characteristics could be engineered through careful control of absorption coefficient and transition temperature.

In particular, better progression of primary liquid front, higher LHSC (%) and liquid fraction for TNP-OC (ST) case is observed in comparison to TNP-OC (GT), NP-OC and TC irrespective of the magnitude of the incident flux and boundary condition. Interestingly, thermal charging route has an edge in the initial phase of melting, however, as the melting progresses, thermochromism assisted optical charging takes the lead owing to increased penetration depth of photons and increased magnitude of nanoparticle-photon interactions. Moreover, for low values of absorption coefficient ($K_{abs}$ = 110 m$^{-1}$ in the present work); a secondary melting front also appears at the bottom - which further enhances the LHSC. It may be noted that this secondary melting front is unique to the TNP-OC route and is not observed in other charging routes. Finally, the most significant advantage of the TNP-OC charging routes is their ability to significantly accelerate the melting rate under nearly thermostatic conditions. In other words, melting rate could be significantly enhanced without appreciably raising the temperature of the PCM.

**ACKNOWLEDGEMENTS**

V.K. acknowledges the support provided by DST-SERB (under sanction order no. CRG/2021/003272). V. K. also acknowledges the support provided by Mechanical Engineering Department, Thapar Institute of Engineering & Technology Patiala, India. IS acknowledges the support provided by Mechanical Engineering Department, Chandigarh University, Gharuan.

**Appendix A1.**

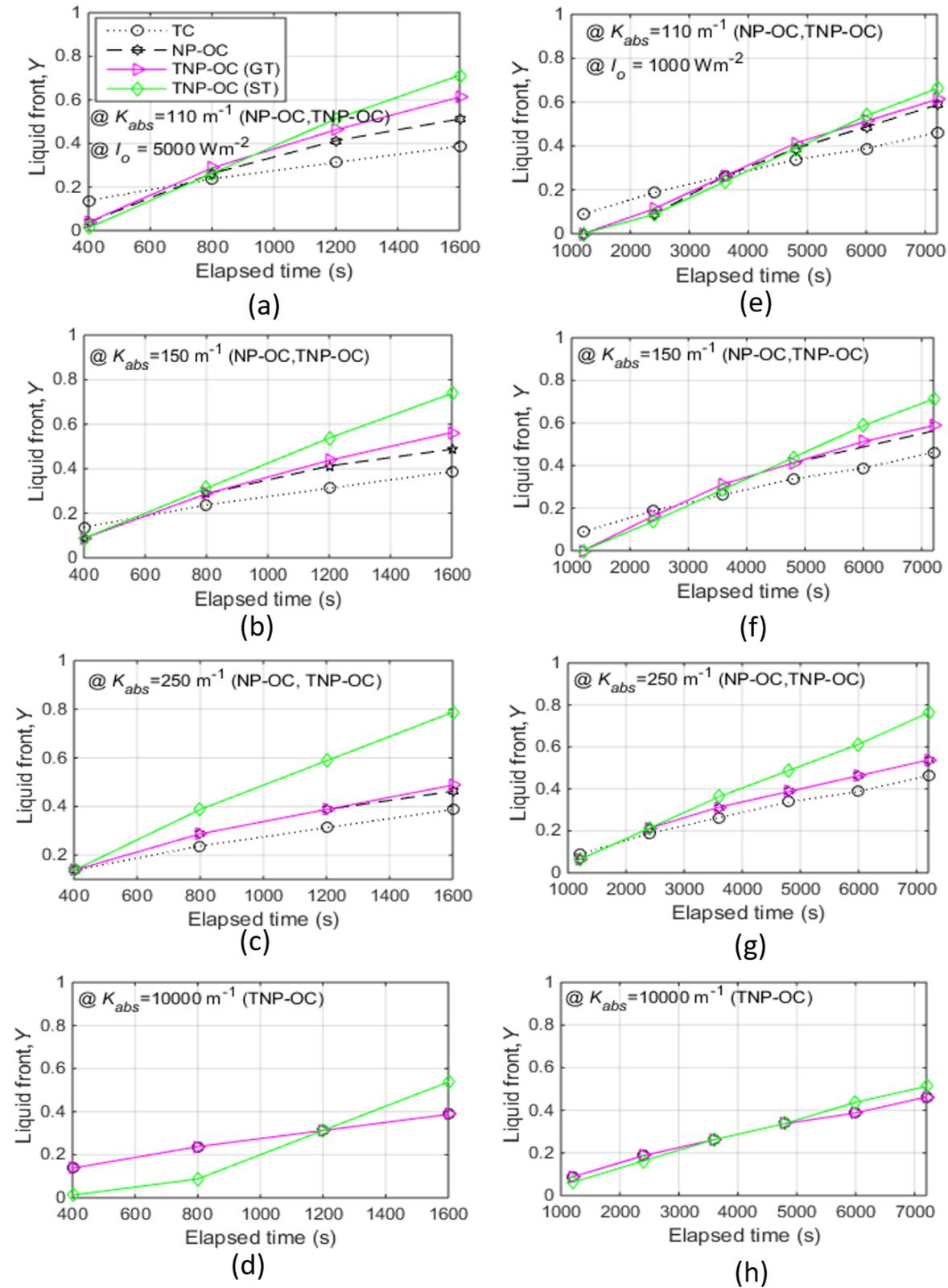

Fig. A1 Mid-plane comparison of primary liquid front position for nano-PCM, TC ($K_{abs}$ = 10000 m$^{-1}$) and OC ( viz. NP-OC, TNP-OC (GT), TNP-OC (ST))  corresponding to flux $I_o$ = 5000 Wm$^{-2}$ [(a)  $K_{abs}$ = 110 m$^{-1}$ (b) $K_{abs}$ = 150 m$^{-1}$ (c) $K_{abs}$ = 250 m$^{-1}$ (d)  $K_{abs}$ = 10000 m$^{-1}$]; $I_o$ = 1000 Wm$^{-2}$ [(e) $K_{abs}$ = 110 m$^{-1}$ (f) $K_{abs}$ = 150 m$^{-1}$ (g) $K_{abs}$ = 250 m$^{-1}$ (h) $K_{abs}$ = 10000 m$^{-1}$.

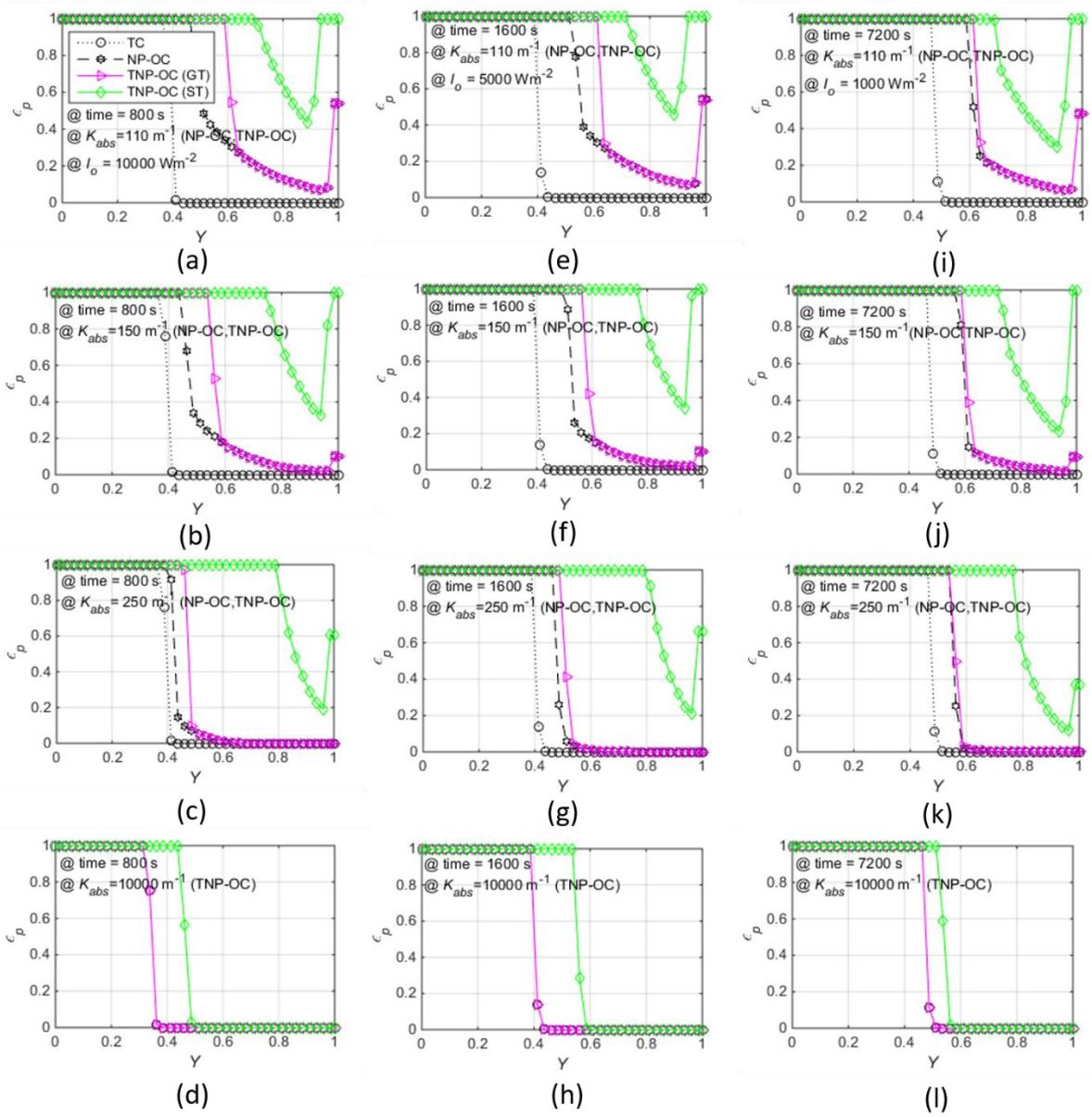

Fig. A2 Comparison of mid-plane liquid fraction for nano-PCM, TC (Kabs =10000 m$^{-1}$) and OC ( viz. NP-OC, TNP-OC (GT), TNP-OC (ST)) corresponding to $I_o$ = 10000 Wm$^{-2}$, time = 800 s [(a) $K_{abs}$ = 110 m$^{-1}$ (b) $K_{abs}$ = 150 m$^{-1}$ (c) $K_{abs}$ = 250 m$^{-1}$ (d) $K_{abs}$ = 10000 m$^{-1}$]; $I_o$ = 5000 Wm$^{-2}$, time = 1600 s [(e) $K_{abs}$ = 110 m$^{-1}$ (f) $K_{abs}$ = 150 m$^{-1}$ (g) $K_{abs}$ = 250 m$^{-1}$ (h) $K_{abs}$ = 10000 m$^{-1}$]; $I_o$ = 10000 Wm$^{-2}$, time = 7200s [(i) $K_{abs}$ = 110 m$^{-1}$ (j) $K_{abs}$ = 150 m$^{-1}$ (k) $K_{abs}$ = 250 m$^{-1}$ (l) $K_{abs}$ = 10000 m$^{-1}$].

**NOMENCLATURE**

*English Symbols:*

| | | | |
|---|---|---|---|
| $A$ | coefficients for each control volume and its faces | $\varepsilon$ | *liquid fraction* |
| $a$ | surface area [m$^2$] | $\upsilon$ | kinematic viscosity of nanofluid [m$^2$ s$^{-1}$] |
| $B$ | width of enclosure [m] | $\rho$ | density of nanofluid [kg m$^{-3}$] |
| $C_{ps}$ | specific heat [J kg$^{-1}$ K$^{-1}$] | $\sigma$ | Stefan Boltzmann constant [Wm$^{-2}$k$^{-4}$] |
| $D$ | depth of enclosure [m] | $\tau$ | glass transmissivity |
| $g$ | acceleration due to gravity [ms$^{-2}$] | *Subscripts:* | |
| $H$ | enthalpy | *abs* | absorption |
| $h$ | convection heat transfer coefficient [Wm$^{-2}$K$^{-1}$] | *amb* | ambient |
| $I$ | Incident solar flux [Wm$^{-2}$] | $E$ | control volume east of $P$ |
| $K$ | absorption coefficient [m$^{-1}$] | $e$ | east face of control volume |
| $k$ | thermal conductivity [Wm$^{-1}$K$^{-1}$] | $N$ | control volume north of $P$ |
| $L$ | length of enclosure [m] | $n$ | north face of control volume |
| $P$ | pressure | *ref* | Reference |
| $Pr$ | Prandtl number | $S$ | control volume south of $P$ |
| $Q''$ | total convective and radiative losses | $s$ | south face of control volume |
| $T$ | local fluid temperature [K] | $s$ | south face of control volume |
| $T_p$ | mid-plane temperature [K] | $B$ | bottom surface |
| $u$ | velocity component of fluid along x-direction [ms$^{-1}$] | $L$ | left wall surface |
| $v$ | velocity component of fluid along y- direction [ms$^{-1}$] | $R$ | right wall surface |
| $x$ | horizontal coordinate direction [m] | *t_conv* | top convective loss |
| $y$ | vertical coordinate direction [m] | *t_rad* | top radiative loss |
| $X$ | dimensionless width of enclosure, $x/L$ | *visc* | viscosity |
| $Y$ | dimensionless depth of enclosure, $y/H$ | *Superscript* | |
| $dx$ | thickness of each CV along x direction | $T$ | term depicting temperature variable |
| $dy$ | thickness of each CV along y direction | $H$ | term depicting enthalpy variable |
| | | *Abbreviations* | |
| *Greek symbols*: | | TC | thermal charging |
| $\alpha$ | thermal diffusivity [m$^2$s$^{-1}$] | NP-OC | non thermochromic-nanoparticle optical charging |
| $\beta$ | coefficient of volumetric expansion [K$^{-1}$] | TNP-OC | thermochromic nanoparticle based optical charging |
| $\varepsilon_g$ | emissivity of glass plate [= 0.88] | GT | gradual transition |
| $\Omega$ | relaxation factor for enthalpy updation | ST | sudden transition |
| $\mu$ | dynamic viscosity [Kgm$^{-1}$s$^{-1}$] | | |
| $\omega$ | under-relaxation factor for transport equations | | |
| $\Omega$ | relaxation factor for enthalpy updation | | |
| $K$ | spectral index of absorption | | |
| $\mu$ | dynamic viscosity [kgm$^{-1}$s$^{-1}$] | | |
| $\Delta$ | change in latent heat | | |